          [2001/04/25 1.1 (PWD)]

\RequirePackage{fix-cm}
\documentclass[smallextended]{svjour3}       
\smartqed  
\usepackage{times}
\usepackage{graphicx}
\usepackage{mathptmx}      
\begin{document}

\title{In-flight measurement of Planck telescope emissivity}

\author{F. Cuttaia [1]      
\and L. Terenzi [1]
\and G. Morgante [1]
\and M. Sandri [1]
\and F. Villa [1]
\and A. De Rosa [1]
\and E. Franceschi [1]
\and M. Frailis [2]
\and S. Galeotta [2]
\and A. Gregorio [2]
\and P. Delannoy [3]
\and S. Foley [3]
\and B. Gandolfo [3]
\and A. Neto [3]
\and C. Watson [3]
\and F. Pajot [4]
\and M. Bersanelli [5]
\and R. ~C. Butler [1]
\and N. Mandolesi [1,6]
\and A. Mennella [5]
\and J. Tauber [7]
\and A. Zacchei [2]
}


\institute{Francesco Cuttaia \at
              INAF/OAS - Osservatorio di Astrofisica e Scienza dello Spazio di Bologna, Via Piero Gobetti 101, 40129, Bologna, Italy \\
              Tel.: +39 051 6398771\\
              \email{francesco.cuttaia@inaf.it}           
\and \\
\and[1] INAF - Osservatorio di Astrofisica e Scienza dello Spazio di Bologna, Via Gobetti 101, 40129, Bologna, Italy\\
\and[2] INAF - Osservatorio Astronomico di Trieste, Via G.B. Tiepolo 11, 34143, Trieste, Italy\\
\and[3] Mission Operations Centre (MOC), ESA/European Space Operations Centre, Robert-Bosch-Str. 5, 64293 Darmstadt, Germany \\
\and[4] Institut de Recherche en Astrophysique et Planétologie 9, avenue du Colonel Roche - BP 44346, 31028 Toulouse cedex 4, France \\
\and[5] Dip. di Fisica, Universit\`a degli Studi di Milano, Via Celoria 16, 20133, Milano, Italy\\
\and[6] Dip. di Fisica, Universit\`a degli Studi di Ferrara,Via Saragat 1 44122 - Ferrara,Italy\\
\and[7] Science Directorate, European Space Agency, Keplerlaan 1, 2201AZ Norrdwijk, The Netherlands}

\maketitle

\begin{abstract}
The Planck satellite in orbit mission ended in October 2013. Between the end of Low Frequency Instrument (LFI) routine mission operations and the satellite decommissioning, a dedicated test was also performed to measure the Planck telescope emissivity.\\
The scope of the test was twofold: i) to provide, for the first time in flight, a direct measure of the telescope emissivity; and ii) to evaluate the possible degradation of the emissivity by comparing data taken in flight at the end of mission with those taken during the ground telescope characterization.\\ 
The emissivity was determined by heating the Planck telescope and disentangling the system temperature excess measured by the LFI radiometers.\\
Results show End of Life (EOL) performance in good agreement with the results from the ground optical tests and from \textit{in-flight} indirect estimations measured during the Commissioning and Performance Verification (CPV) phase.\\
Methods and results are presented and discussed.
\end{abstract}

\section{Introduction}
\label{sec:introduction}
The Planck satellite \cite{Tauber2010},\cite{Planck_coll_1_2011} was launched together with the Herschel spacecraft on an Ariane 5 from Europe's spaceport in Kourou, French Guyana, on 14 May 2009. The two satellites were injected into an orbit around the Sun-Earth Lagrange point $L_2$. The duration of the nominal Planck mission was 15,5 months. Nevertheless Planck operated continuously for 1623 days, until 23  October 2013, with the Low Frequency Instrument (LFI). The High Frequency Instrument (HFI) \cite{Lamarre2010} operated until 13 January 2012 when the supply of $^3$He needed to cool the HFI bolometers to 0.1 K ran out. However, the HFI's He Joule-Thomson cooler \cite{Planck_coll_2_2011} 
continued to operate normally to support the LFI pseudocorrelation radiometers \cite{Bersanelli2010} with the required thermal reference at $\sim$4K \cite{Valenziano2011} until mission end.\\
The Planck de-orbiting started on 14 August 2013, when the first manoeuvre for the spacecraft departure from L2 was performed: this phase lasted until October 9th (final de-orbiting manoeuvre).\\ 
In the period between October 4th and October 21st, the LFI functionality at End of Life (EOL) was verified: some tests, already performed during the CPV phase \cite{Gregorio2014} or during the ground calibration tests \cite{Mennella2011}, \cite{Terenzi2010}, \cite{Cuttaia2010}, were repeated. New additional tests were also performed to verify or better characterize other features revealed during the mission.\\
In particular, the procedure named Telescope Loss Test (TLT)  was run: it was aimed at indirectly measuring the Planck telescope emissivity \cite{Tauber_2_2010} at the LFI frequencies at EOL, by operating the de-contamination heaters located on the primary and the secondary mirrors. This test was not foreseen at the beginning of the Planck mission and was decided upon only during the planning of Planck EOL phase, taking advantage of the LFI radiometers sensitivity \cite{Mennella_2_2011} and of our improved knowledge of the LFI properties and of systematic effects over the mission \cite{Planck_coll_2014}, \cite{Planck_coll_2015}.\\
The test consisted in heating the primary and secondary mirrors by few Kelvin ($\sim 4K$) and then measuring the power excess measured by the LFI pseudo-correlation radiometers. The underlying basic assumption was that the measured excess would be mostly proportional to the telescope reflection loss (emissivity), provided that the other possible effects affecting the radiometric response were known and kept under control.\\ 
The TLT procedure was succesfully run on 7 October 2013. Such a test was never performed before on a microwave space telescope: the high LFI instrumental sensitivity, a very good knowledge of systematic effects and the Planck Mission Operation Center (MOC) ability in controlling the telescope thermal response, were the key ingredients of its success.

\section{The Planck Telescope}
\label{sec:Planck_Telescope}
\begin{figure}
	\centering
		\includegraphics [width=1.0\textwidth]{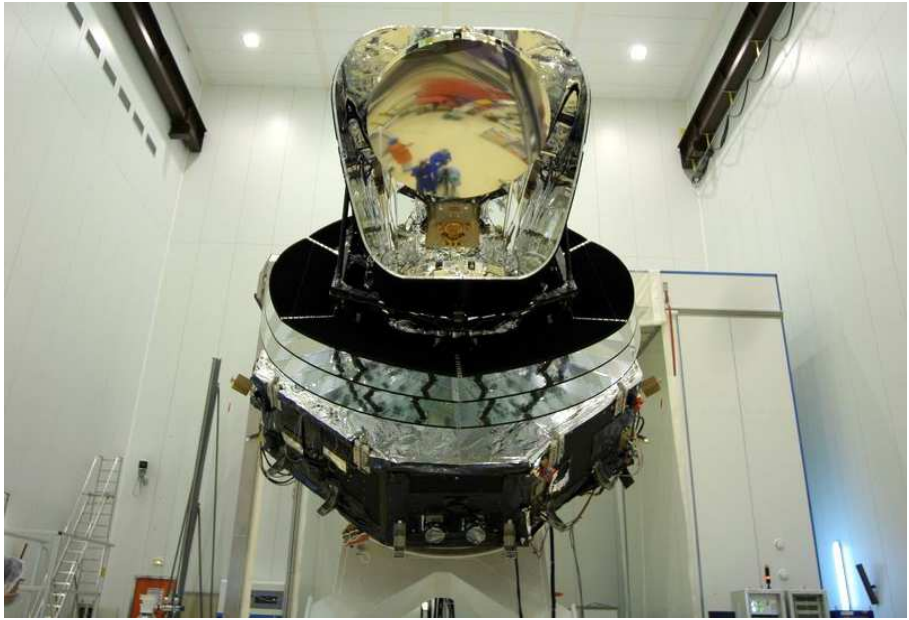}
	\caption{the Planck Satellite, on mounting fixture, during the ground tests. The telescope (primary mirror) and the thermal baffle are visible in the foreground}.
	\label{fig:telescope}
\end{figure}

he Planck telescope was designed to comply with the following high level opto-mechanical requirements: 
\begin{itemize}
	\item wide frequency coverage: about two decades, from 25 GHz to 1 THz;
	\item 100 squared degrees of field of view, wide focal region (400 X 600 mm);
	\item cryogenic operational environment between 40 K and 65 K.
	\end{itemize}
The telescope optical layout was based on a dual reflector off-axis Gregorian design (Fig.~\ref{fig:telescope}). Both the primary and secondary mirrors were elliptical in shape. The size of the primary mirror rim was 1.9 X 1.5 meters; the rim of the secondary mirror was nearly circular with a diameter of about 1 meter.\\

\begin{figure}[h]
 \begin{minipage}[b]{5.8cm}
   \centering
   \includegraphics[width=5.75cm]{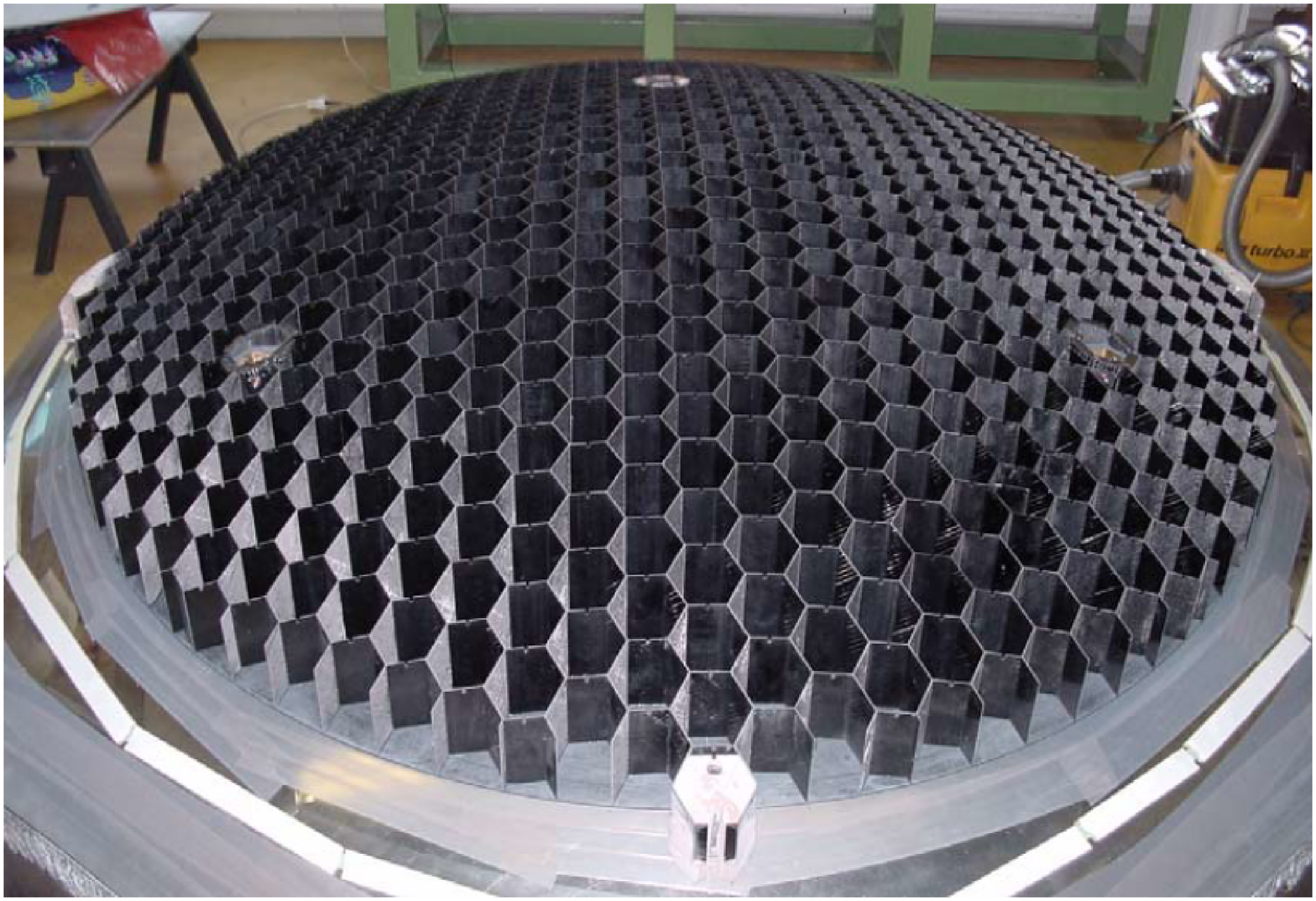}
   \caption{Primary Reflector: milled core }
   	\label{fig:PR_honeycomb}
 \end{minipage}
  \ \hspace{1mm} \hspace{1mm} 
 \begin{minipage}[b]{5.5cm}
  \centering
   \includegraphics[width=5.3cm]{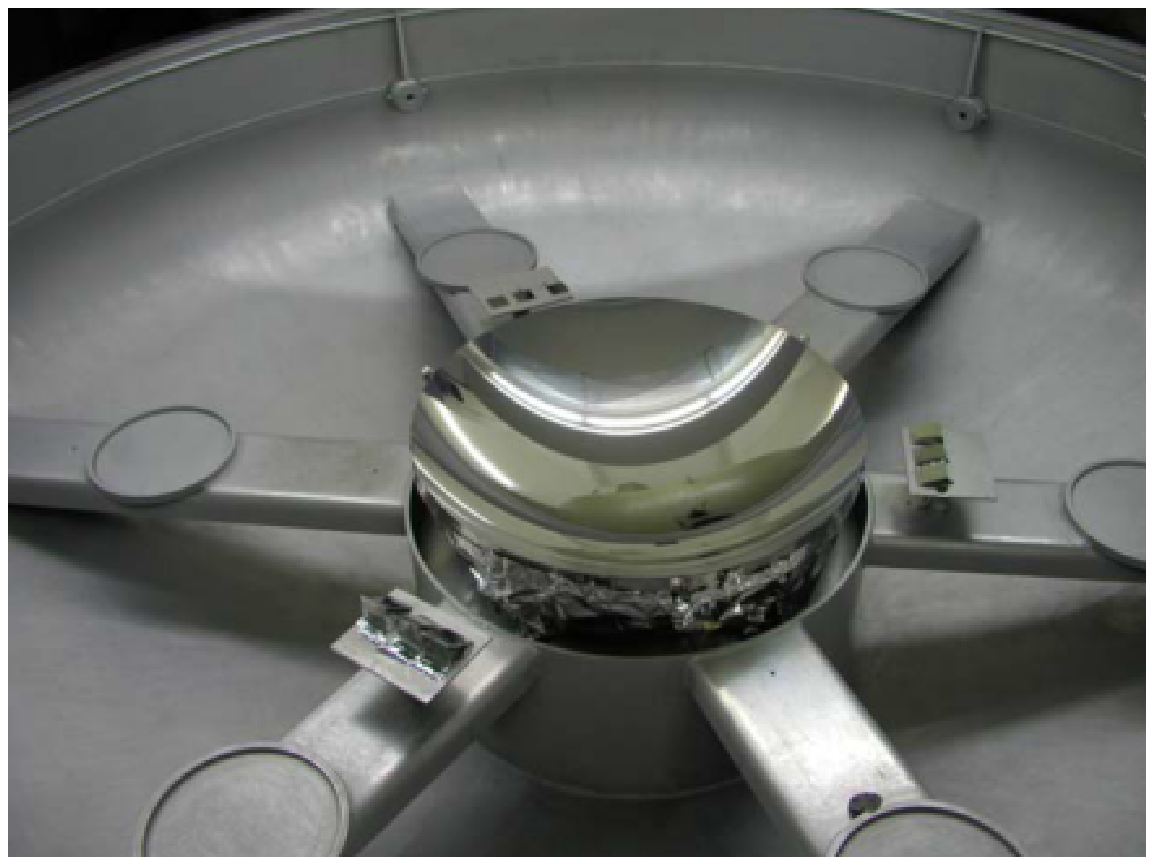}
   \caption{Secondary Reflector: reflective coating}
   	\label{fig:coating_SR}
 \end{minipage}
\end{figure}

The overall focal ratio was 1.1, and the projected aperture was circular with a diameter of 1.5 meters. The telescope field of view was $\pm5 ^{\circ}$ centred on the line of sight (LOS), which was tilted at about $3.7^{\circ}$ relative to the main reflector axis, and formed an angle of $85^{\circ}$ with the satellite spin axis, which was typically oriented in the anti-Sun direction during the survey.\\
The Gregorian off-axis configuration ensured a small overall focal ratio (and thus small feeds), an unobstructed field of view, and low diffraction effects from the secondary reflector and struts.\\ 
The core of primary and secondary mirrors was fabricated using Carbon Fiber Reinforced Plastic (CFRP) honeycomb sandwich technology (Fig.~\ref{fig:PR_honeycomb}). The facesheets underwent reflective coating (Fig.~\ref{fig:coating_SR}), following a procedure developed by EADS Astrium, consisting of three layers: ~15 nm NiCr as adhesion layer, 550 nm Aluminium as reflective layer, $\sim$30 nm PLASIL as protection layer \cite{stute2005}.\\
This design was chosen to satisfy the requirements of low mass ( < 120 Kg including struts and supports), high stiffness, high dimensional accuracy, and low thermal expansion coefficient. 
Further details on the Planck optical system can be found in \cite{Planck_coll_2_2011} and in \cite{stute2004}.

\section{The Telescope Loss Test}
\label{sec:Telescope_Loss}
\begin{figure}[h]
 \begin{minipage}[b]{5.8cm}
   \centering
   \includegraphics[width=5.75cm]{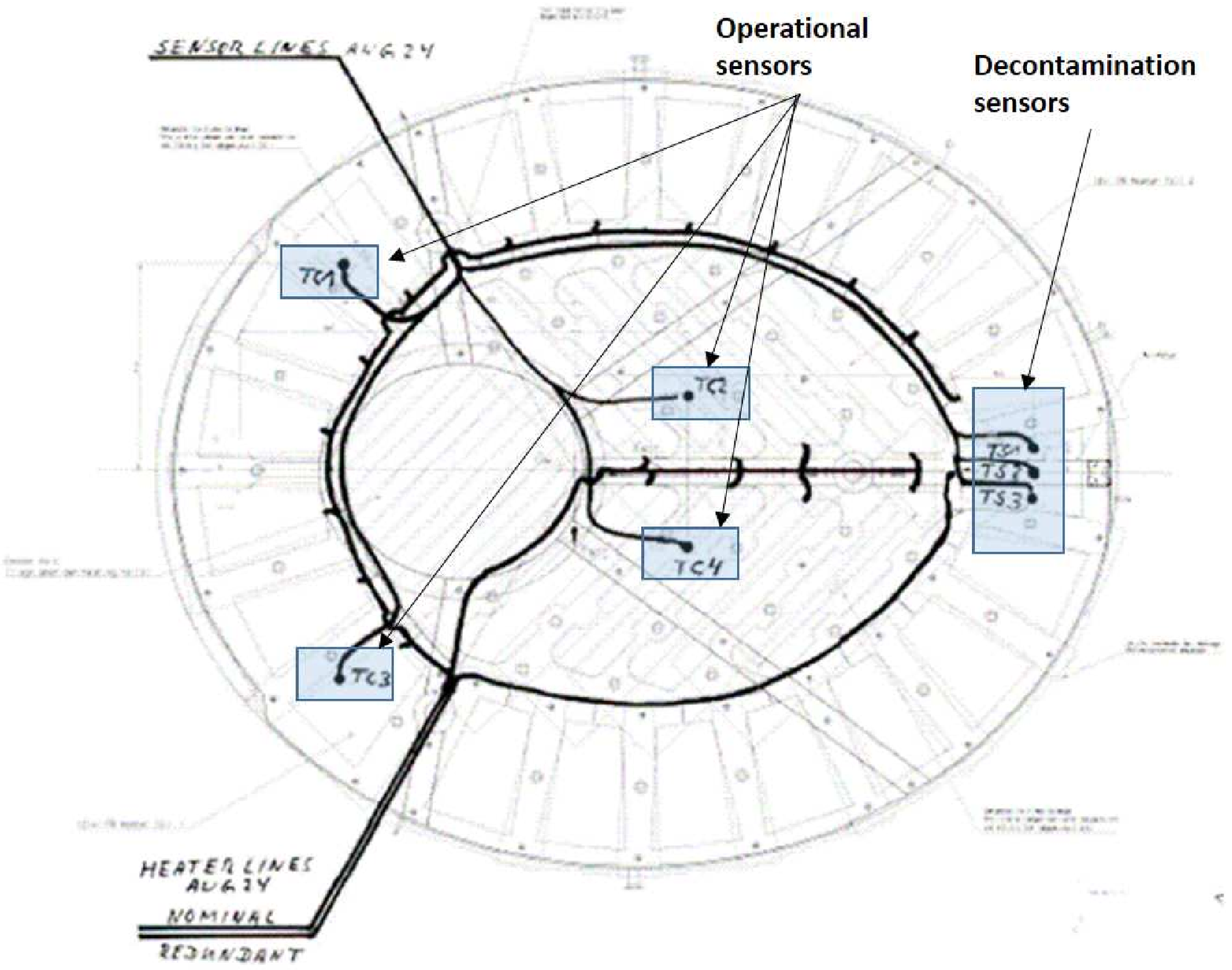}
   \caption{Primary Reflector: heaters harness and temperature sensors location scheme.}
   	\label{fig:sensor_PR}
 \end{minipage}
  \ \hspace{1mm} \hspace{1mm} 
 \begin{minipage}[b]{5.5cm}
  \centering
   \includegraphics[width=5.3cm]{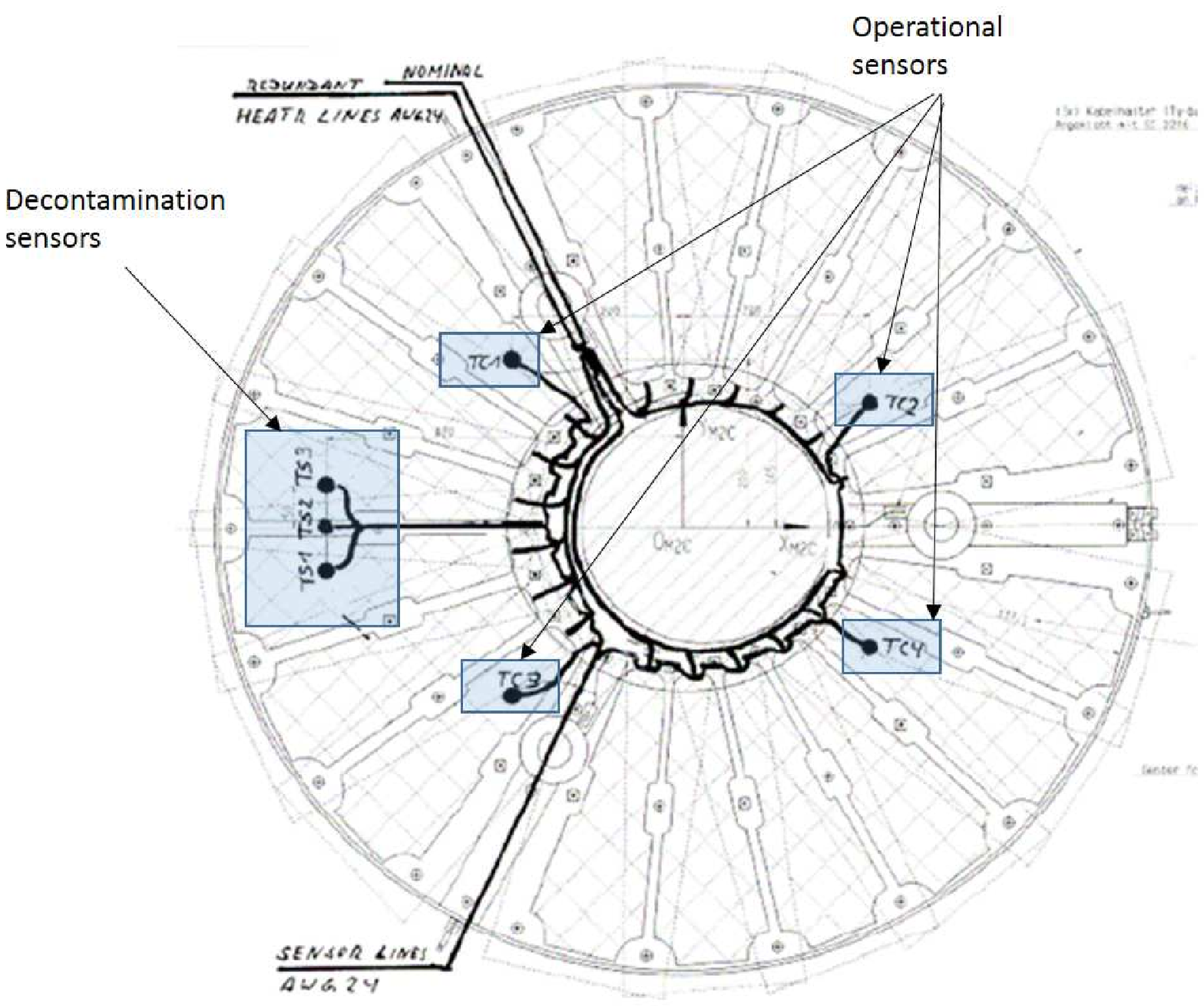}
   \caption{Secondary Reflector: heaters harness and temperature sensors location scheme.}
   	\label{fig:sensor_SR}
 \end{minipage}
\end{figure}

The TLT started on 7 October 2013 at 19:25:00 UTC, when \textit{anti-contamination} heating was activated through heaters placed on the primary (PR) and secondary (SR) reflectors. The heaters were operated adapting to the test, in a cyclic fashion, the algorithm that was originally designed for de-contaminating the reflectors during the early launch phases.\\
Temperatures for decontamination were monitored in real time by three dedicated sensors for each of the Planck reflectors (the three adjacent sensors in line in figs. ~\ref{fig:sensor_PR} and ~\ref{fig:sensor_SR}), while temperatures used for analysis are measured with a better resolution (about 0.25 K istead of 0.5 K) by two couples of nominal and redundant sensors (the four symmetrically distributed sensors in figs. ~\ref{fig:sensor_PR} and ~\ref{fig:sensor_SR}), for each reflector.\\

On the  basis of the emissivity measured during the ground tests (to be lower than  0.0006 at the LFI frequency) a minimum temperature change of 2K was required to unambiguously characterize the in-flight emissivity: nevertheless, the overall decontamination procedure was able to increase the temperature of the PR and SR by roughly 4K (averaged over the corresponding monitoring sensors), while the temperatures remained quite stable for the last 90 minutes of the test.\\ Finally, both reflectors started to cooldown at a rate of less than 1K in 12 hours.\\
Temperature profiles of PR and SR, caused by anti-contamination heaters activation and de-activation, are respectively shown in Fig.~\ref{primary_solo} and in Fig.~\ref{secondary}.

\section{Emissivity Characterization}
\label{sec:Emissivity}
\begin{figure}[htbp]
	\centering
		\includegraphics[width=0.80\textwidth]{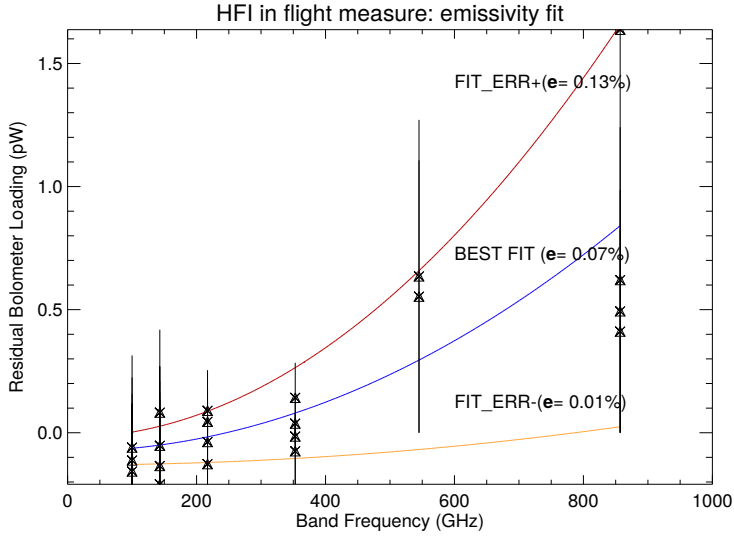}
	\caption{Emissivity fit calculated by HFI from thermal arguments. The plot displays the Residual Bolometer Loading (pW) versus Frequency (GHz). All frequency channels are evenly constraining the emissivity of the mirrors (a common error on the bolometer plate temperature thermometer is considered): this leads to an estimate of 0.07 +/- 0.06\% for each of the two mirrors, using a Rayleigh-Jeans law. Three cases are shown: best fit (0.07\%, blue line), fit with positive error (0.07 + 0.06\%, red line), fit with negative error (0.07 - 0.06\%, orange line). Error bars correspond to experimental errors at each frequency for each detector. The best fit, and the two uncertainty curves, result from considering all points simultaneously.}
	\label{fig:planck_telescope_emissivity_w_error}
\end{figure}

The emissivity plays a crucial role in microwave telescopes, even more in spinning telescopes like Planck. Actually, the black-body thermal emission from the telescope is the cause of a higher system temperature; moreover, thermal fluctuations of the telescope can mimic the effect of changes in sky emission, critical expecially at fluctuation frequencies near the satellite spin frequency. For this reason the telescope emissivity was required, at beginning of life (BOL), to be lower than 0.6\% .\\
The telescope emissivity was expected to change during the mission due to UV irradiation and micrometeoroid impact, especially at the HFI frequencies.\\
The emissivity was estimated on ground, by measuring the reflection loss of several samples from the Herschel telescope \cite{Fisher2005}. However, due to non-neglible differences between the Herschel and Planck telescopes, more accurate tests, based on a high-quality open Fabry-Perot resonator, were performed in 2008,directly on same Planck telescope samples, between 100 GHz and 380 GHz \cite{parshin2008}, at the Institute of Applied Physics of the Russian Academy of Sciences (IAP RAS).  Measures performed at low temperature (between 80 K and 110 K), showed that the reflectivity of mirror surfaces basically depends on: i) the quality of thin reflecting metal layers, ii) the coating, iii) the temperature . Results show an emissivity lower than the requirement, in the frequency range  100 GHz- 380 GHz.\\
The emissivity was also  measured indirectly in flight by the HFI, from thermal arguments: the background power in the bolometer bands, coming from the primary and secondary mirrors, was measured for each detector. Results, reported in \cite{Planck_coll_A2_2011} show an emissivity of about 0.07\%, an order of magnitude lower than the requirement, obtained from the least squares fit of the computed in-band power from the two mirrors: emissivity is assumed to be frequency independent. As reported in  \cite{Planck_coll_A2_2011}, these results are affected by a large uncertainty (up to 100\%) , especially at the two highest frequencies (545 GHz and 857 GHz), possibly due to calibration error in the bolometer plate temperature thermometer or thermal gradients between thermometer and bolometers location (Fig.~\ref{fig:planck_telescope_emissivity_w_error}).\\

The TLT allowed to measure emissivity also in the Planck complementary frequency range covered by LFI radiometers. To first order, the mean differential power output for each of the four receiver diodes of the LFI radiometers can be written as (Eq.1, \cite{Mennella_2_2011}):

   \begin{equation}
        P_{\rm out}^{\rm diode} = a\, G_{\rm tot}\,k\,\beta \left[ \tilde{T}_{\rm sky} + T_{\rm noise} - r\left(
            T_{\rm ref} + T_{\rm noise}\right) \right],
        \label{eq_p0}
    \end{equation}
  
    where $G_{\rm tot}$ is the total gain, $k$ is the Boltzmann constant, $\beta$ the receiver bandwidth and $a$ is the detector constant. $\tilde{T}_{\rm sky}$ and $T_{\rm ref}$ are, respectively, the apparent average sky antenna temperature and the reference load antenna temperature at the inputs of the first hybrid; $T_{\rm noise}$ is the receiver noise temperature. $\tilde{T}_{\rm sky}$ is the apparent sky signal entering the first hybrid after the two reflections on the primary and on the secondary mirros. The two reflections combine, attenuating the true sky signal and adding a  spurious thermal signal proportional to the emissivity of the mirrors.  
    The gain modulation factor (Eq.2, \cite{Mennella_2_2011}), $r$, is defined by:
    \begin{equation}
        r = \frac{\tilde{T}_{\rm sky} + T_{\rm noise}}{T_{\rm ref} + T_{\rm noise}},
        \label{eq_r}
    \end{equation}

In order to accurately characterize the telescope emissivity a good knowledge of the following quantities is mandatory:
\begin{itemize}
	\item \textsc{PR and SR temperatures}: they affect $T_{\rm sky}$. To this aim, we must consider that the thermal sensors have limited resolution (about 0.2K).\\
	\item \textsc{4K Reference Load (4KRL) stage thermal stability}: it affects $T_{\rm ref}$. Instabilities at 4KRL  level impact on the differenced output of LFI radiometers, mimicking a change in the measured sky signal.\\
	\item \textsc{LFI detectors calibration constants}: they affect $G_{\rm tot}$. Any error in the calibration constants propagates as a multiplicative error in the emissivity.\\ 
	\item \textsc{Front End Unit (FEU) thermal stability}: it affects $G_{\rm tot}$. Instabilities at FEU level impact on the gain of the front end low noise amplifiers.\\  
	\item \textsc{Back End Unit (BEU) thermal stability}: it affects $G_{\rm tot}$ and $a$. Instabilities at BEU level can either impact on the gain and bias offset of the back end low noise amplifiers or on the radiometers power suppliers (controlling the FEU LNAs gain) or on both.\\
\end{itemize}

A detailed analysis of these systematic effects is given in \cite{Planck_coll_2014}, \cite{Planck_coll_2015}.\\
All these systematic effects were accounted for in the test preparation and execution and in the data analysis.\\
The total signal transmitted from PR ans SR can be written as:

\begin{eqnarray}
T_{\rm PR}^{\rm out} \approx (1-\epsilon_{\rm 1})T_{\rm sky} + \epsilon_{\rm 1}T_{\rm PR}\\
T_{\rm SR}^{\rm out} \approx (1-\epsilon_{\rm 2})[(1-\epsilon_{\rm 1})T_{\rm sky}+\epsilon_{\rm 1}T_{\rm PR}]+\epsilon_{\rm 2}T_{\rm SR}\\
T_{\rm SR}^{\rm out}\equiv \tilde{T}_{\rm sky}
  \label{eq:eq_T_reflectors}       
\end{eqnarray}

PR and SR were manufacured following a common procedure and using the same materials. For this reason, we can assume that:

\begin{equation}
\epsilon_{\rm 1} \approx \epsilon_{\rm 2} \sim \epsilon
\end{equation}

Reducing the above equations, we get a simple expression relating the antenna temperature variation to thermal excess due to reflectors heating:

\begin{equation}
\Delta T^{\rm ant} \approx \epsilon (\Delta T_{\rm PR}+\Delta T_{\rm SR})
\end{equation}

\section{Data Analysis}
\label{sec:Data_Analysis}
The differenced output from each diode of the LFI detectors was correlated to the nominal temperature changes of the primary and secondary reflectors. The temperature associated to the reflectors was the average among the sensors respectively monitoring the PR and the SR.\\
Data were calibrated averaging nominal gains calculated during one day in the late routine phase (day 1480 after launch) before the TLT. Calibration constant used are reported in the Appendix (Tab.~\ref{cal_const}).\\
The effect of signal fluctuations induced by the dipole modulation, caused by the Planck Telescope spinning, was also taken into account. Results, after dipole contribution removal, differ only negligibly from those before correction.\\
The results were also corrected for radiometer susceptibility to temperature changes of: the front end unit (\textsc{FEU}), the back end enit (\textsc{BEU}) and the 4K stage (\textsc{4KRL}). Also with respect to these systematic effects, differences were negligible, because of the high thermal stability of the LFI during the TLT test.\\ 

\begin{figure}[h!]
    \begin{center}
        \includegraphics[width=8 cm]{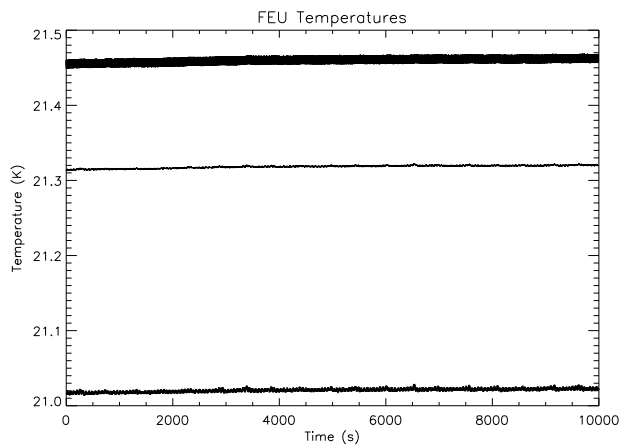}\\
    \end{center}
    \caption{Fron end unit sensors positioned near feedhorn LFI28, LFI25, LFI26. They correspond to the three LFI Q band channels (\cite{Bersanelli2010})}
    \label{FEU_temp} 
\end{figure}

\begin{figure}[h!]
    \begin{center}
        \includegraphics[width=8 cm]{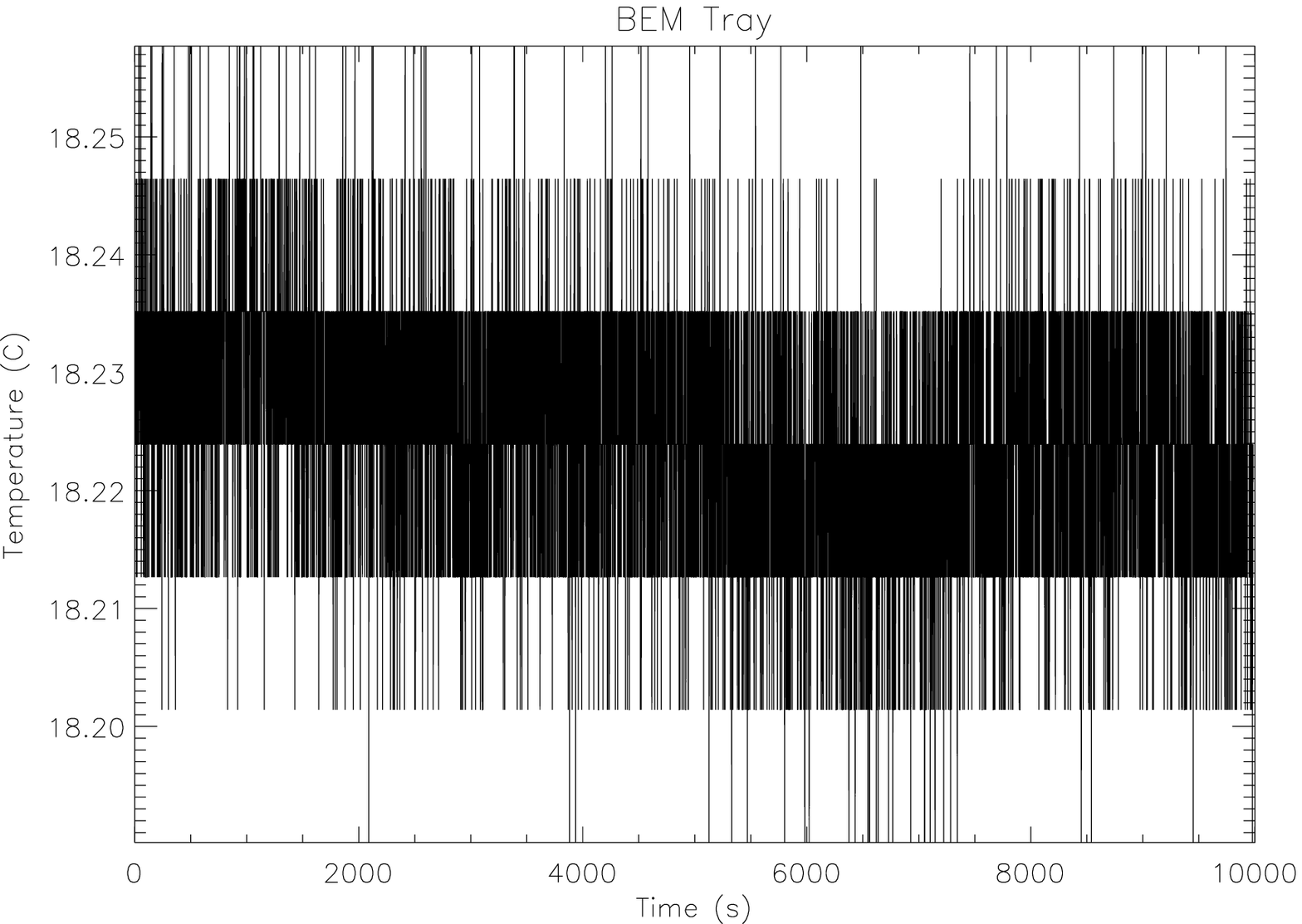}\\
    \end{center}
    \caption{Back end unit sensors positioned on BEM tray (\cite{Bersanelli2010})}
    \label{BEM_tray} 
\end{figure}

\begin{figure}[h!]
    \begin{center}
        \includegraphics[width=8 cm]{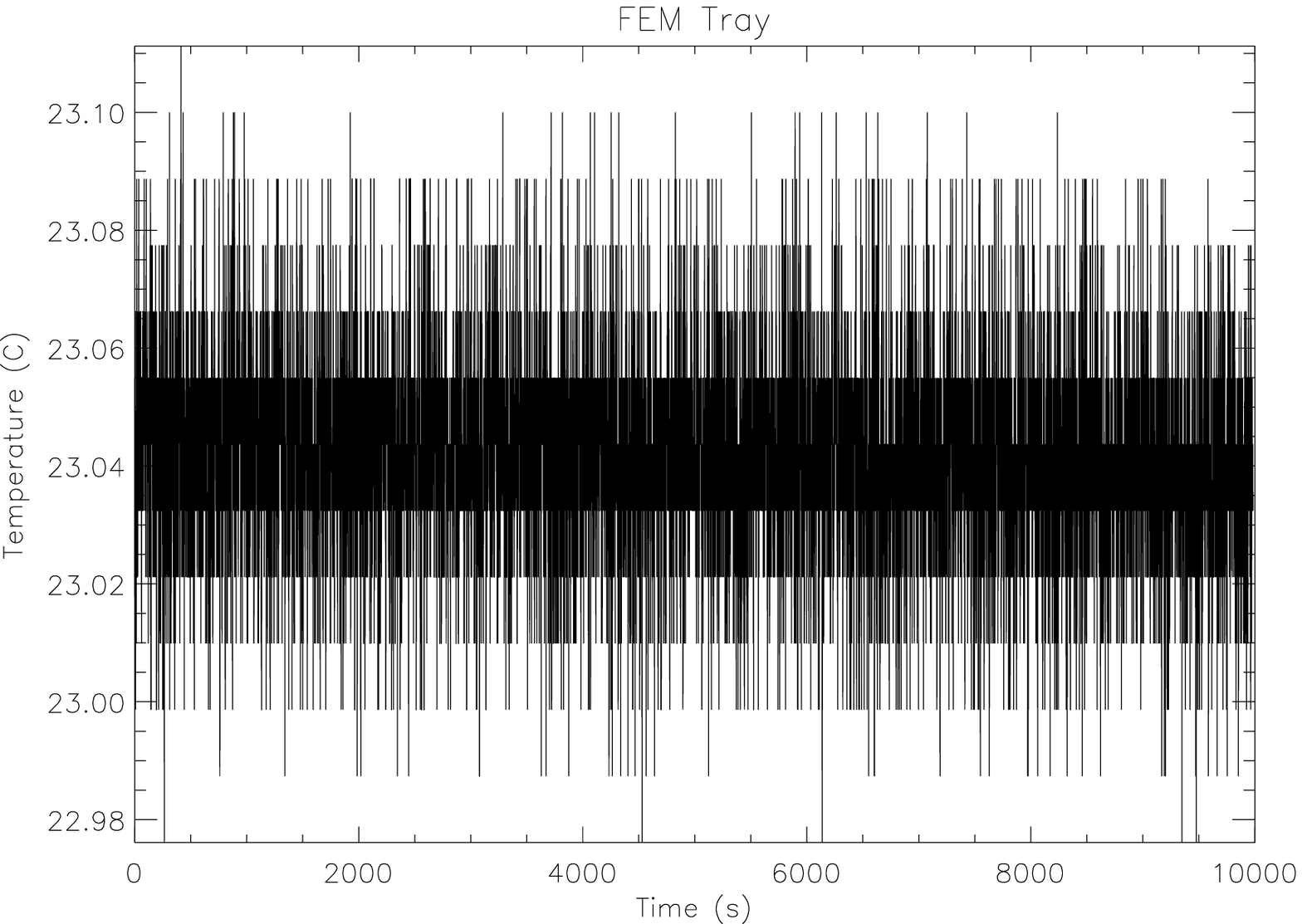}\\
    \end{center}
    \caption{Back End Unit sensors positioned on FEM tray (\cite{Bersanelli2010})}
    \label{FEM_tray} 
\end{figure}

\begin{figure}[h!]
    \begin{center}
        \includegraphics[width=8 cm]{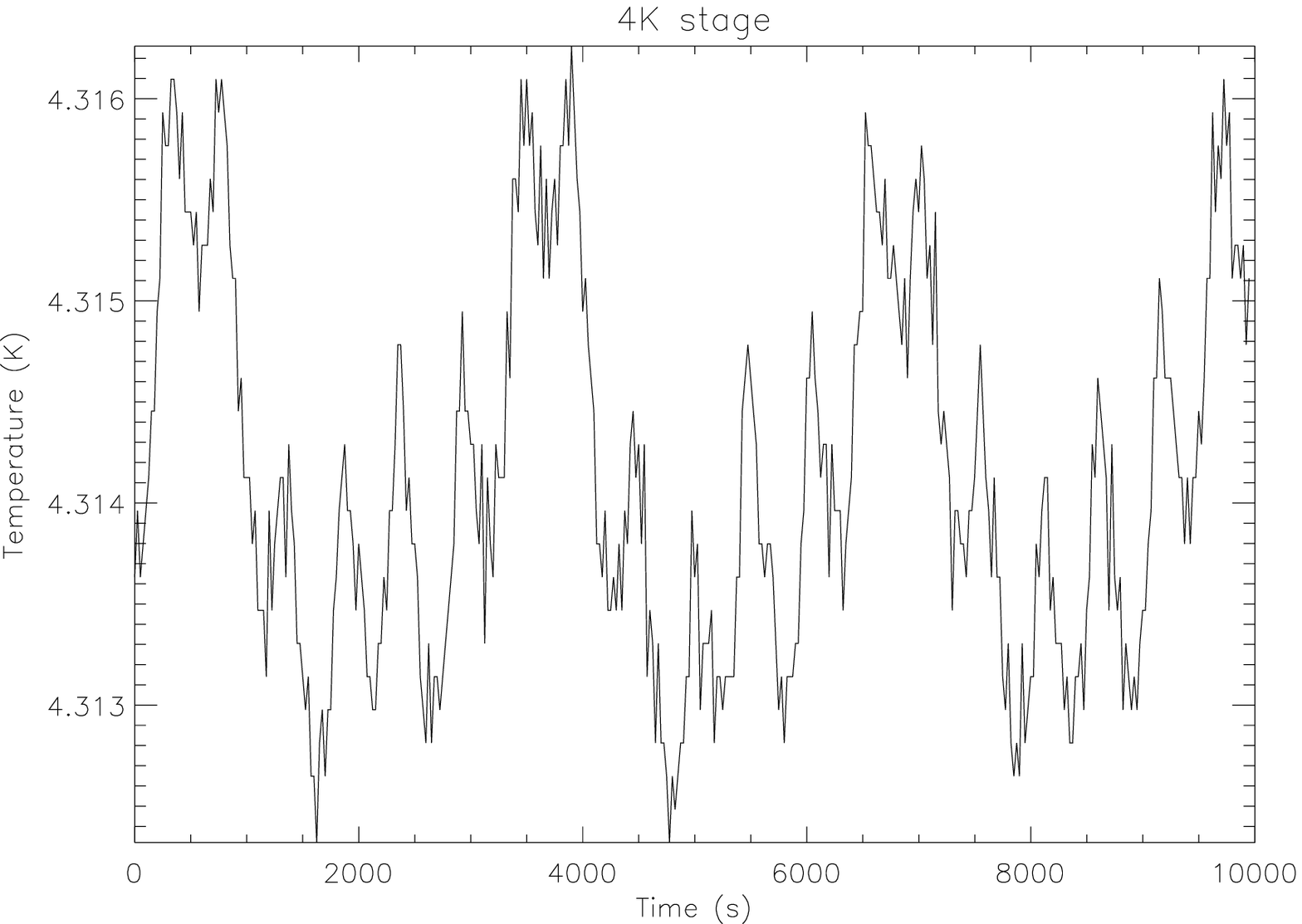}\\
    \end{center}
    \caption{4K stage temperature}
    \label{4K} 
\end{figure}

The LFI thermal behavior is shown in the following figures. Peak to peak variations are:
\begin{itemize}
	\item lower than 0.05 K in the \textsc{FEU} (Fig.~\ref{FEU_temp} );
	\item at the level of sensors resolution in the BEU (Fig.~\ref{BEM_tray} and Fig.~\ref{FEM_tray} show quantized signals ); 
	\item lower than 4 mK in the 4K Reference Load Unit (Fig.~\ref{4K});
\end{itemize}
All the above effects do not show any correlations with temperature changes in PR and SR. Reference values for the thermal susceptibilities are those from \cite{Planck_coll_2014}.\\

\begin{figure}[h!]
    \begin{center}
        \includegraphics[width=8 cm]{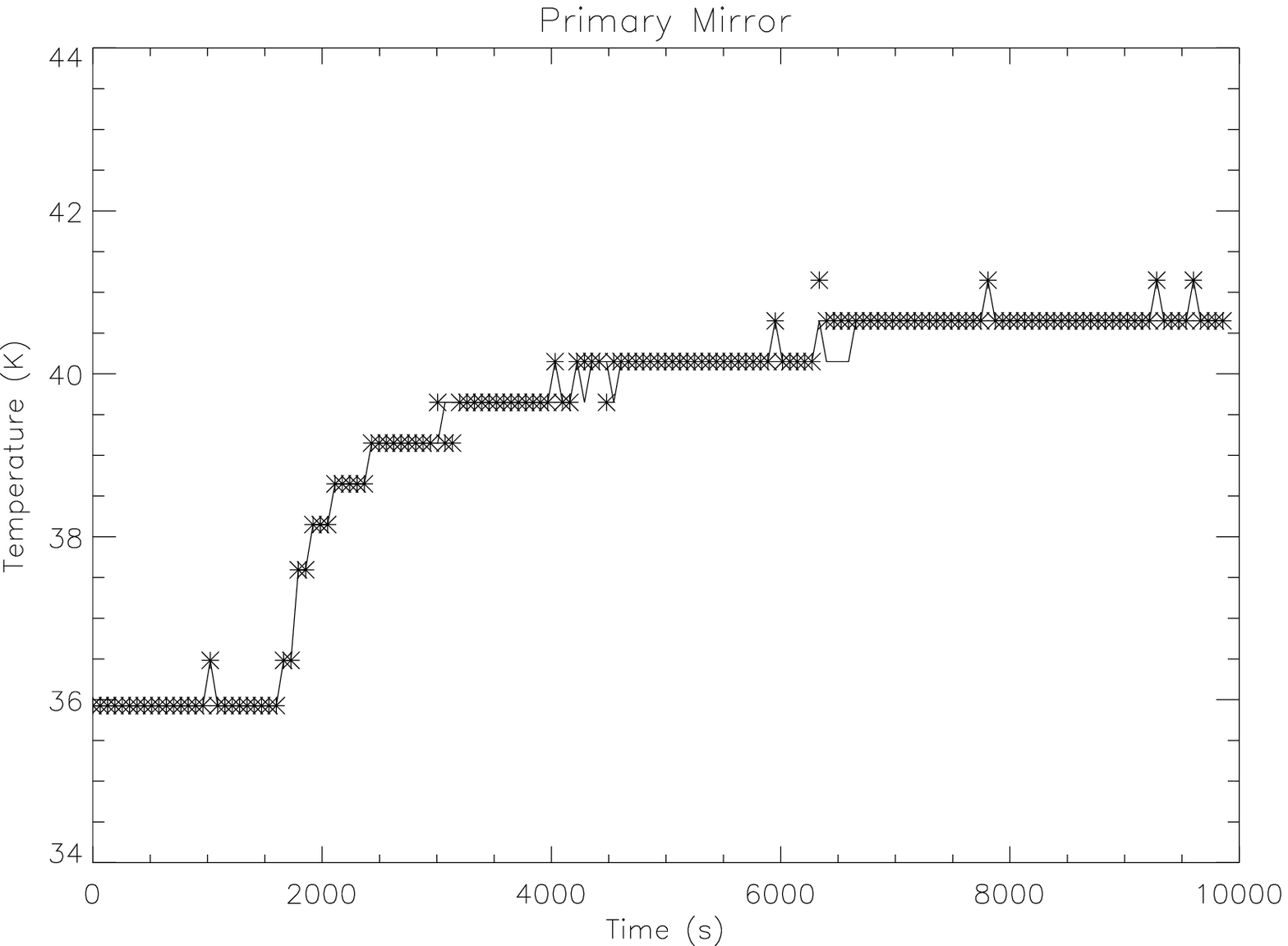}\\
    \end{center}
    \caption{Primary reflector temperature averaged between three PR sensors}
    \label{primary_solo} 
\end{figure}

\begin{figure}[h!]
    \begin{center}
        \includegraphics[width=8 cm]{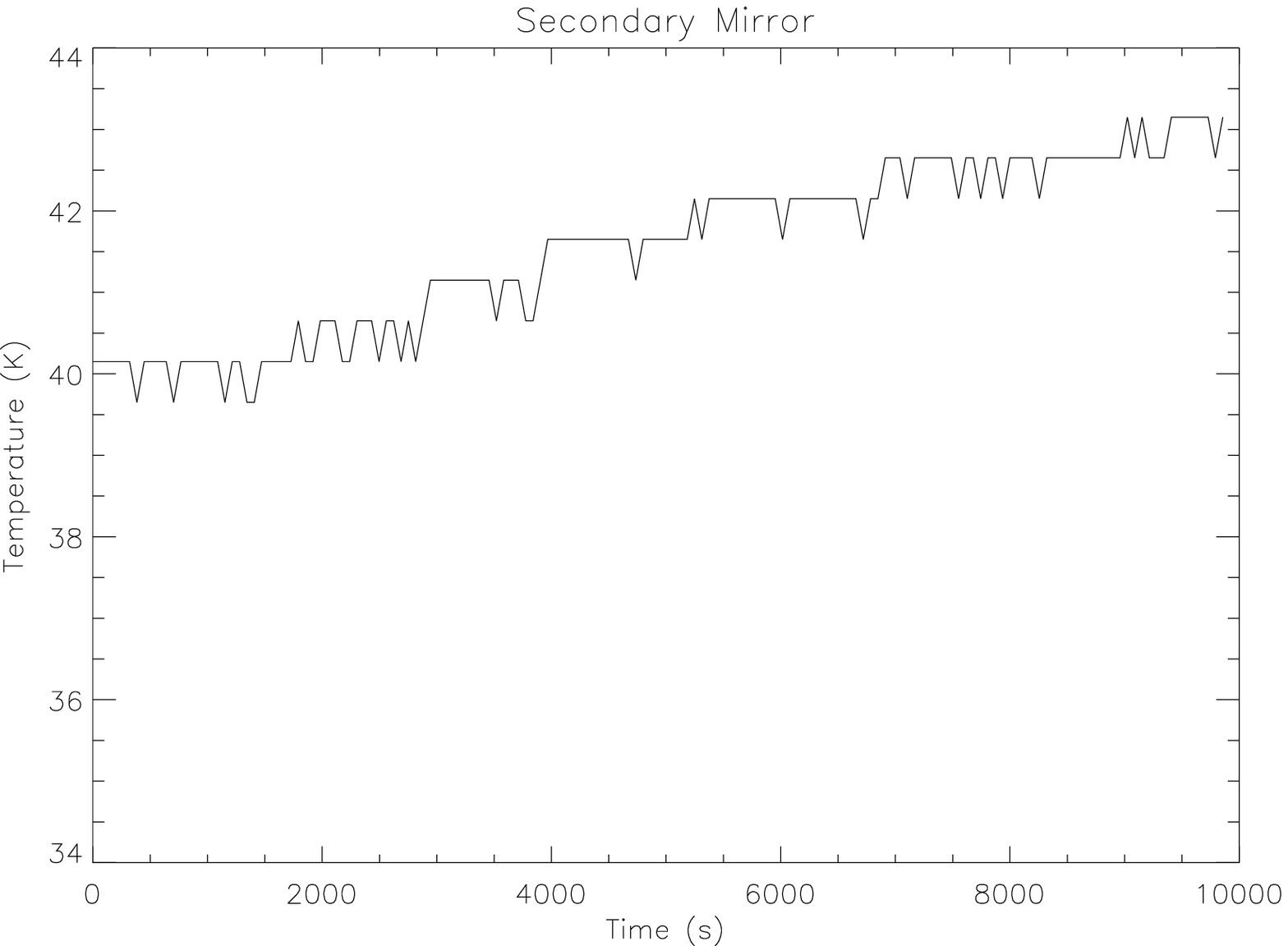}\\
    \end{center}
    \caption{Secondary reflector temperature averaged between three SR sensors}
    \label{secondary} 
\end{figure}

The temperature variation of primary and secondary reflectors, averaged over the sensors monitoring each reflector, is displayed in Fig.~\ref{primary_solo} and in Fig.~\ref{secondary}. The relevant quantity is not the absolute temperature, but instead the thermal change due to reflector heating.\\ 

\begin{figure}[h!]
    \begin{center}
        \includegraphics[width=8 cm]{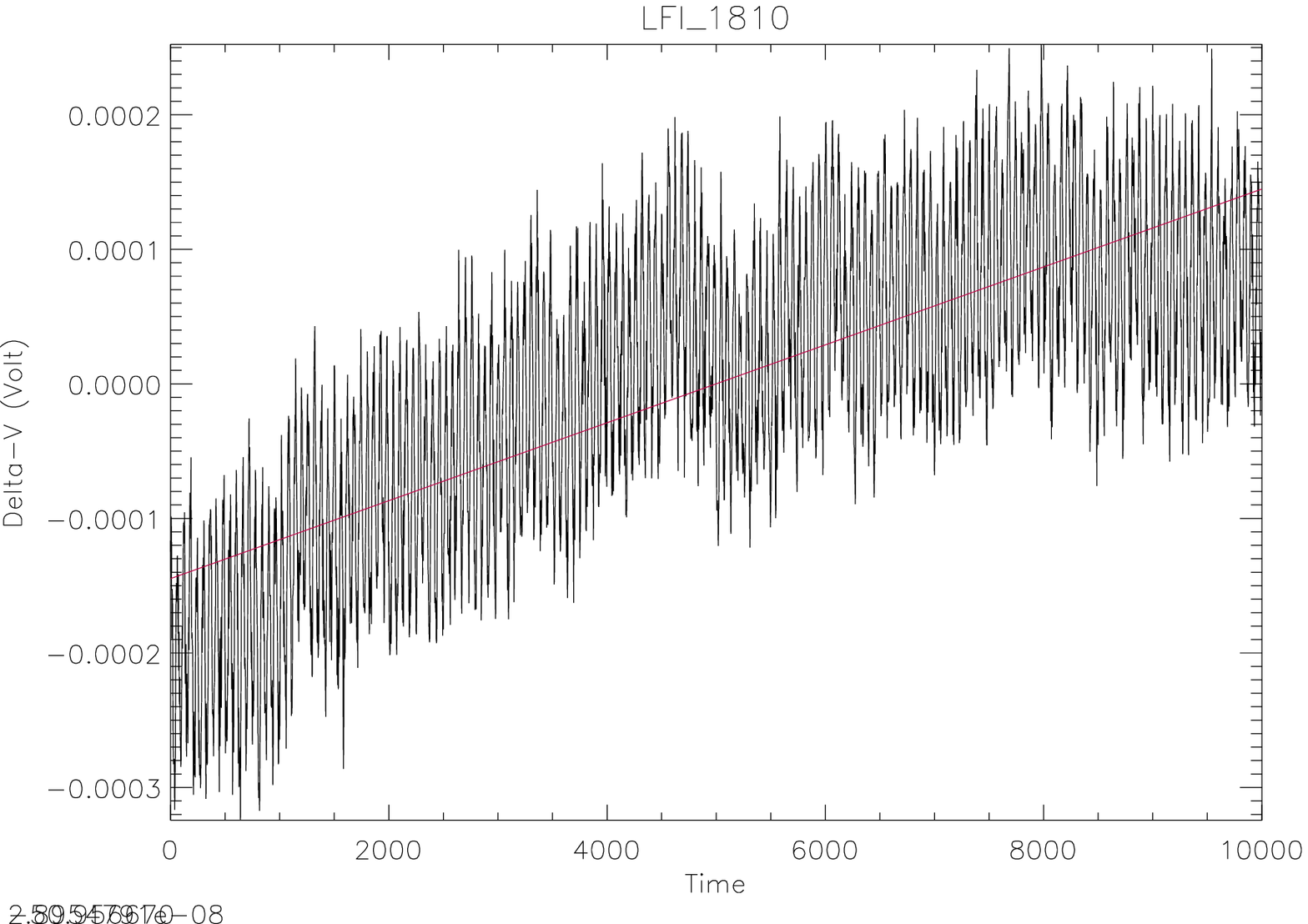}\\
    \end{center}
    \caption{differenced output change in the 70 GHz channel LFI-1810 caused by telescope heating during TLT}
    \label{RCA1810_reb} 
\end{figure}

\begin{figure}[h!]
    \begin{center}
        \includegraphics[width=8 cm]{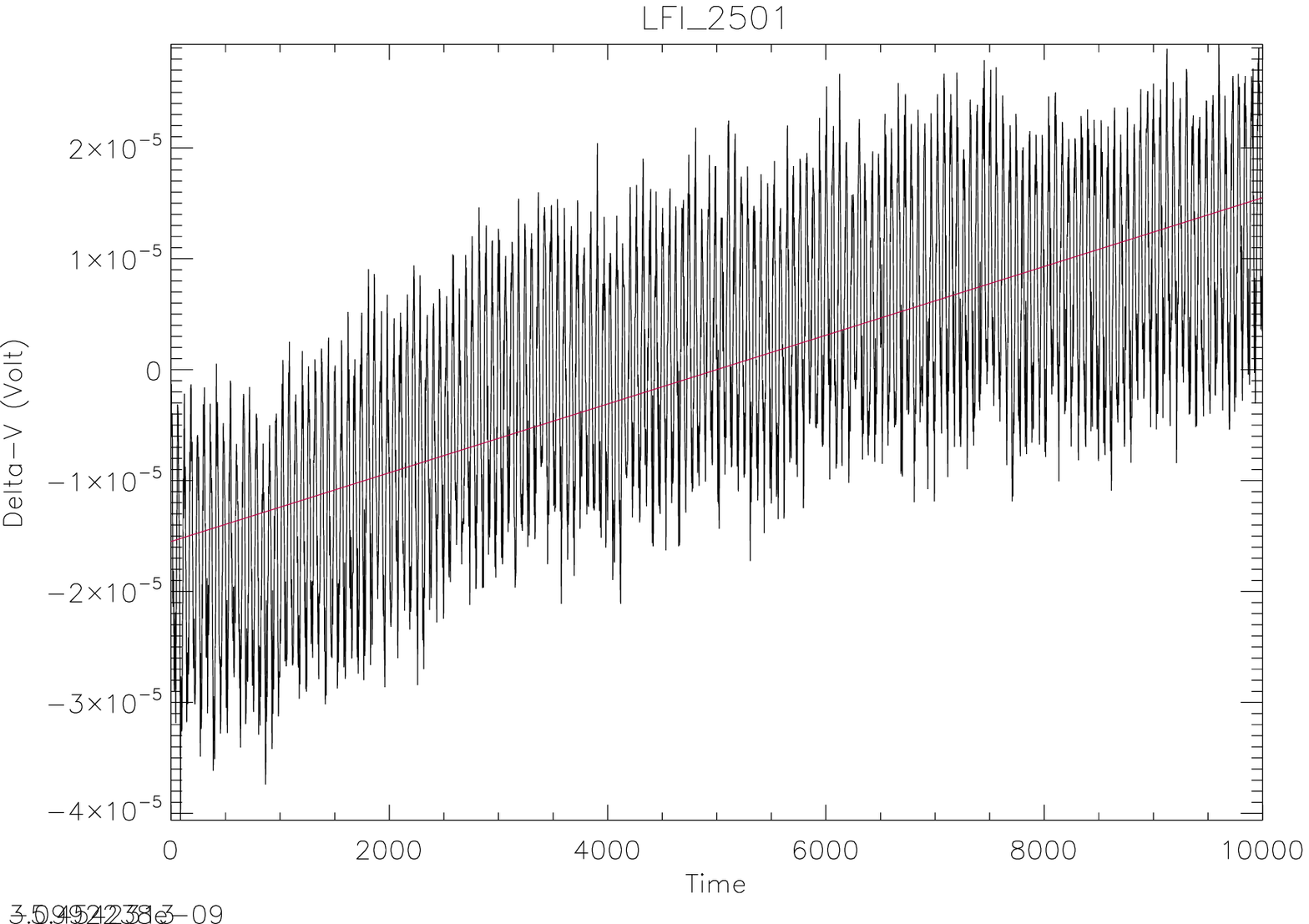}\\
    \end{center}
    \caption{differenced output change in the 44 GHz channel LFI-2501 caused by telescope heating during TLT}
    \label{RCA2501_reb} 
\end{figure}

\begin{figure}[h!]
    \begin{center}
        \includegraphics[width=8 cm]{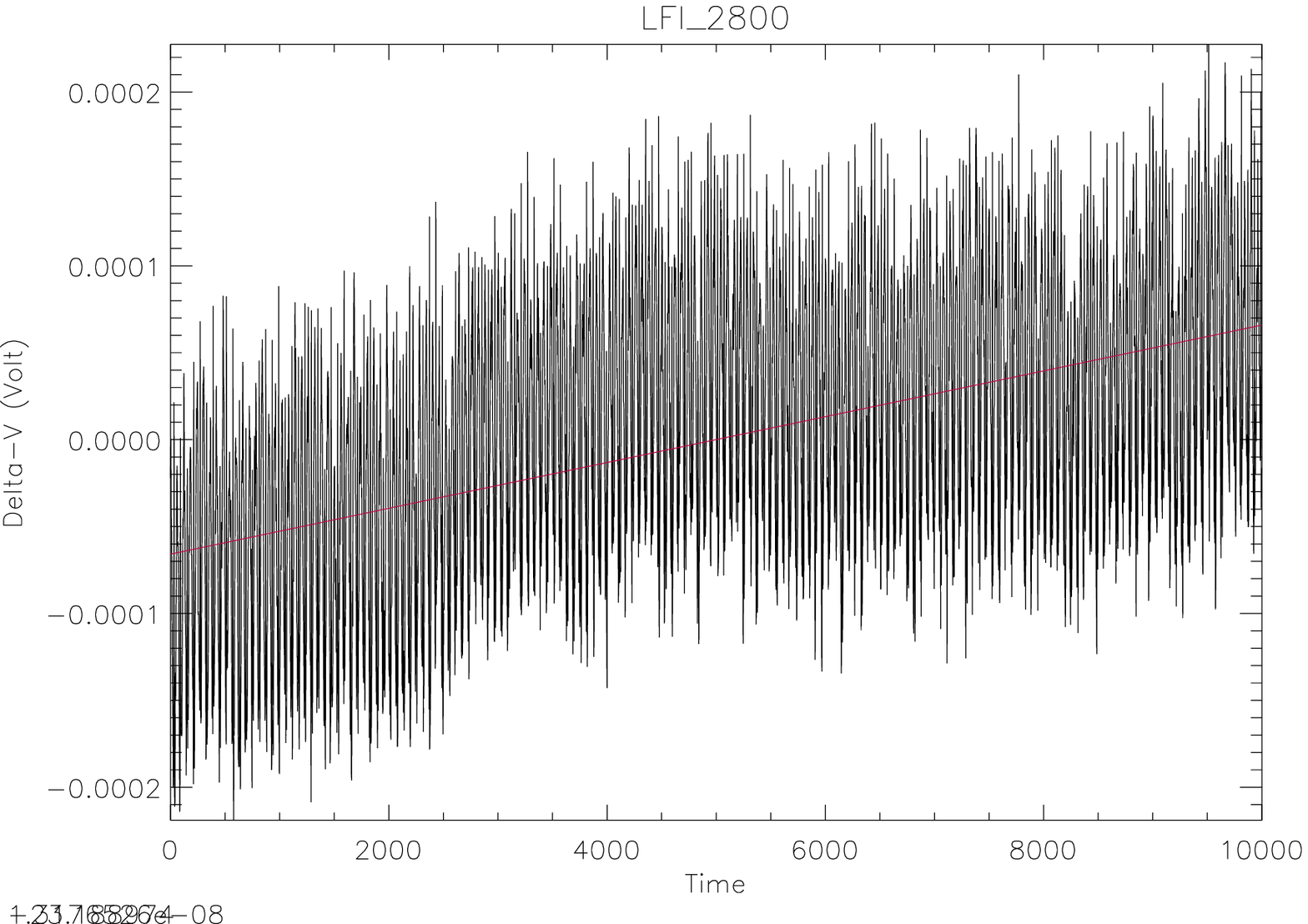}\\
    \end{center}
    \caption{differenced output change in the 30 GHz channel LFI-2800 caused by telescope heating during TLT}
    \label{RCA2800_reb} 
\end{figure}

The effect of PR and SR heating on radiometers is hereafter shown for three channels  \begin{ttfamily} LFI18, LFI25, LFI28\end{ttfamily}, representative of the full LFI frequency range (70 GHz, 44 GHz, 30 GHz respectively), in Fig.~\ref{RCA1810_reb}, Fig.~\ref{RCA2501_reb}, Fig.~\ref{RCA2800_reb}: in order to simplify the visualization, data have been rebinned. The differential nature of the LFI radiometers and their high sensitivity (\cite{Mennella2011}) makes it possible to identify clearly the sky temperature excess due to reflectors heating.\\  

\section{Results}
\label{sec:Results}
Results are presented for each frequency channel in Tab.~\ref{emissivity_frequency}.\\
Results are presented per radiometer (\begin{ttfamily}Main,Side\end{ttfamily})(\cite{Bersanelli2010}) in Tab.~\ref{emissivity_rad}, showing for each channel the emissivity corresponding to the measured apparent sky temperature excess caused by telescope heating. Differenced outputs from coupled diodes are linearly combined as described in \cite{Planck_coll_2_2014}.\\
A different way to combine results is shown in Tab.~\ref{emissivity_channel}, where the measured excess is presented averaging over the LFI optically paired channels (observing the same region of the secondary reflector): this approach is aimed at accounting for possible inhomogeneities in the temperature of the reflectors. Channels have been paired basing on the scheme reported in Tab.~\ref{emissivity_channel} (channel \begin{ttfamily} LFI24\end{ttfamily} is not considered, as it is not paired to any other channels).

\begin{table}[h!]
  \begin{center}
    \caption{Telescope Emissivity. Results are displayed per frequency channel together with the associated uncertainties (calculated as standard deviatiation of emissivities of all radiometers sharing the same frequency}.
    \label{emissivity_frequency}
    \begin{tabular}{l l l}
      \hline
      \hline
	\multicolumn{1}{c}{RCA} & \multicolumn{1}{c}{Emissivity)} & \multicolumn{1}{c}{St.Dev)}\\
	\hline
\texttt{70 GHz}&	5.55E-04 & 1.19E-04	\\
\noalign{\vskip	8pt}	
\texttt{44 GHz}&	4.74E-04 & 8.74E-05	\\
\noalign{\vskip	8pt}	
\texttt{30 GHz}&	3.85E-04 & 5.54E-05 \\
\hline
      \end{tabular}
\end{center}
\end{table}

\begin{table}[ht!]
  \begin{center}
    \caption{Telescope Emissivity. Results are displayed per radiometer. M and S correspond to  \texttt{MAIN} and \texttt{SIDE} radiometers \cite{Bersanelli2010}}.
    \label{emissivity_rad}
    \begin{tabular}{l l l}
      \hline
      \hline
	&\multicolumn{2}{c}{R{\sc adiometer}} \\
	\multicolumn{1}{c}{RCA} & \multicolumn{1}{c}{M}& \multicolumn{1}{c}{S}\\
	\hline
\texttt{LFI18}&	5.25E-04	&	6.19E-04	\\
\texttt{LFI19}&	6.18E-04	&	6.48E-04	\\
\texttt{LFI20}&	5.76E-04	&	6.87E-04	\\
\texttt{LFI21}&	5.51E-04	&	6.15E-04	\\
\texttt{LFI22}&	5.47E-04	&	5.51E-04	\\
\texttt{LFI23}&	2.18E-04	&	5.11E-04	\\
\noalign{\vskip	8pt}			
\texttt{LFI24}&	3.57E-04	&	5.32E-04	\\
\texttt{LFI25}&	4.79E-04	&	4.91E-04	\\
\texttt{LFI26}&	5.93E-04	&	3.93E-04	\\
\noalign{\vskip	8pt}			
\texttt{LFI27}&	3.03E-04	&	3.99E-04	\\
\texttt{LFI28}&	4.20E-04	&	4.18E-04	\\
\hline
      \end{tabular}
\end{center}
\end{table}

\begin{table}[h!]
  \begin{center}
    \caption{Telescope Emissivity. Results are displayed per paired channels, corresponding to feedhorns looking the same region of the telescope. \texttt{MAIN} and \texttt{SIDE} radiometers data have been averaged \cite{Bersanelli2010}.}
    \label{emissivity_channel}
    \begin{tabular}{l l}
      \hline
      \hline
	\multicolumn{1}{c}{RCA} & \multicolumn{1}{c}{Temperature (K)}\\
	\hline
\texttt{LFI18-LFI23}&	4.68E-04	\\
\texttt{LFI19-LFI22}&	5.91E-04	\\
\texttt{LFI20-LFI21}&	6.07E-04	\\
\noalign{\vskip	8pt}	
\texttt{LFI25-LFI26}&	4.89E-04	\\
\noalign{\vskip	8pt}	
\texttt{LFI27-LFI28}&	3.85E-04	\\
\hline
      \end{tabular}
\end{center}
\end{table}

\begin{figure}[h!]
    \begin{center}
        \includegraphics[width=8 cm]{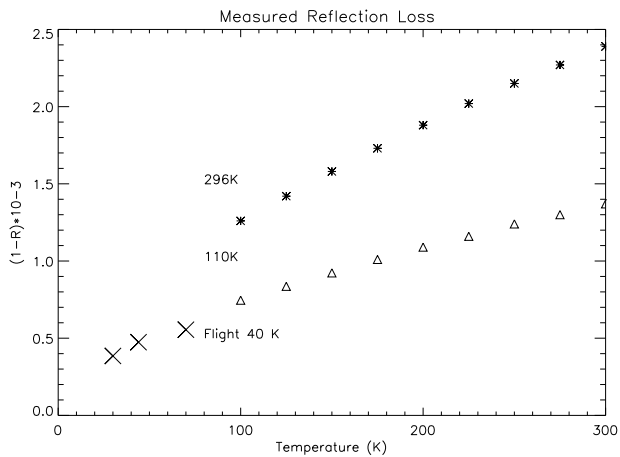}\\
    \end{center}
    \caption{In-flight Reflection Loss at 40K compared to data from\cite{Tauber2010}. They are respectively shown: experimental data from ground test @296 K (asterisc); experimental data from ground test @110 K; Experimental data from LFI in-Flight measurement (this work) at 40K (cross)}
    \label{plot_300_110} 
\end{figure}

The telescope emissivity, per frequency channels (values from Tab.~\ref{emissivity_frequency}), was compared to values reported in Appendix B of \cite{Tauber2010}, where the measured dependence of the Reflection Loss (1-R) of a sample of Planck reflector material is shown at 110 K, as a function of frequency, in the range 100 GHz - 380 GHz. \\
Differences in the Reflection Loss are expected between \textit{in-flight} tests (TLT) and \textit{on-ground} tests (\cite{Tauber2010}), due to the different temperature of the telescope: the largest differences are expected at high frequency \cite{parshin2008}. This frequency dependency allowed to superpose results from TLT test, obtained at 40K in the range 27 GHz - 77 GHz, to results from Fig.B1 -right in \cite{Tauber2010}, obtained at 110 K and at 296 K in the range 100 GHz - 380 GHz. Comparison is shown in Fig.~\ref{plot_300_110} \\

\begin{figure}[h!]
    \begin{center}
        \includegraphics[width=8 cm]{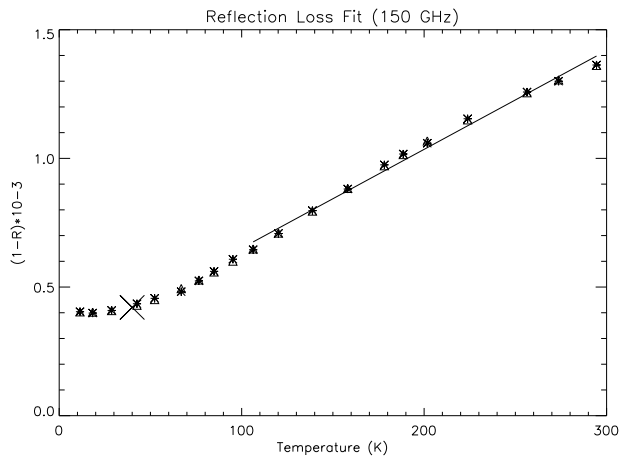}\\
    \end{center}
    \caption{Reflection Loss at from data in \cite{Serov2016}, at 150 GHz. Stars: experimental data. Triangles: forth order polynomial fit; solid line: linear fit in the range [110K : 296K]. Cross: extrapolation down to 40K.}
    \label{plot_extrapolation} 
\end{figure}

\begin{flushleft}
\scshape High Frequency Data Extrapolation to 40K
\end{flushleft}
A more accurate comparison among the data sets, at different temperatures, is obtained by extrapolating to 40K the high frequency data (100 GHz - 380 GHz) measured at 110 K.\\
The temperature dependence of the Reflection Loss was modeled by taking into account the experimental evidences described in \cite{Serov2016}, where the case of mirrors of highly pure aluminum (99.99\% Al) was investigated at fixed frequency f = 150 GHz. When Al is cooled down to cryogenic temperature, experimental results highlight discrepancies with respect to the purely theoretical model, even when anomalous skin depth is considered.  In the case of pure Al, below 150K, the Reflection Loss drops much smoother than predicted, showing an almost linear decrement down to 40K, and constant behaviour at lower temperature (plot 10 in \cite{Serov2016}); the  measured Reflection Loss exceded the theoretical value by about 65\%. Data can be fit with high accuracy (R=0.997) by a 4th order polynomial fit. Nevertheless, also a linear fit in the range 110 K : 300 K is able to predict with good accuracy the behaviour at least down to 40 K. (Fig.~\ref{plot_extrapolation}).

\begin{figure}[h!]
    \begin{center}
        \includegraphics[width=12 cm]{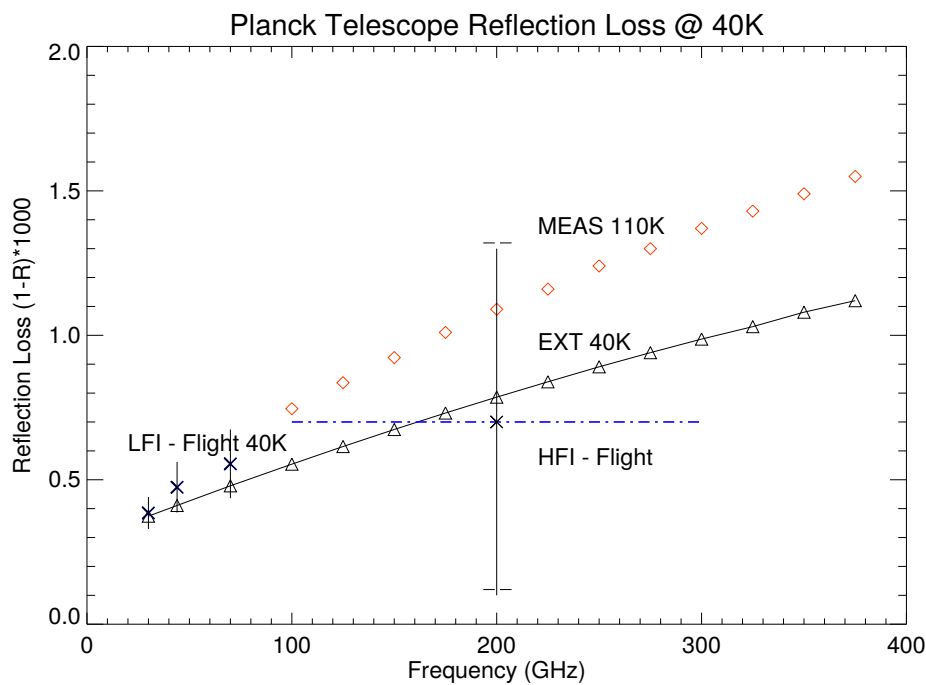}\\
    \end{center}
    \caption{Telescope Reflection Loss. Reflection Loss data are multiplied by a factor of 1000. Results from TLT test, at LFI frequencies 
     are compared to the following data sets: (i) data @110K from Fig.B1 -right in \cite{Tauber2010} (\textsc{MEAS 110K}), red rombs; (ii) data  extrapolated in temperature down to 40K and in frequency down to LFI frequencies (\textsc{EXT 40K}), solid line with triangles; (iii) indirect \textit{in-flight} measure by HFI from thermal arguments (reported in \cite{Planck_coll_2_2011}: \textsc{HFI-Flight}), blue dashed-dot line: the behaviour is flat in frequency; error bars are reported.}
    \label{plot_comparison_40K} 
\end{figure}

At each sampled frequency, the slope was calculated by linearly fitting data in the range 296K-110K; Reflection Loss measured at 110K was hence extrapolated down to 40K. Comparison is presented in Fig.~\ref{plot_comparison_40K}; error bars for data at HFI frequencies, at 296K and 110K, were not available.\\
The emissivity at LFI frequencies - at 40K - is, as expected, lower than the emissivity measured at 110K, at the HFI frequencies, and slightly higher than extrapolated data.\\
Reflection Loss at cryogenic temperature depends on the purity level of the material; in addition, in the specific case of a Space Telescope, cleanliness of the mirrors at the end of mission and aging can play a crucial role. Despite everything, deviations of measured from extrapolated data can be considered negligible, as they are well within the error bars of the in-flight measurement.\\
Results confirm the goodness of this approach and the quality of the Planck telescope, in space - where it was not measured until the end of the mission.


\section{Conclusions}
\label{sec:Conclusions}
The End of Life (EOL)  phase, before Planck satellite de-orbiting, represented a very useful step in completing the characterization of several instrumental properties measured before launch or during the early phases of the mission (CPV).\\ The telescope total emissivity was measured only indirectly (Reflection Loss tests), on test samples of the Herschel telescope first, on samples of the Planck telescope finally. Until EOL, emissivity was measured only in a reduced frequency range covered by HFI (100 GHz -380 GHz), keeping the samples at a temperature (110 K) higher than the in-flight nominal temperature of the telescope (around 40 K).\\ The presence of de-contamination heaters and temperature sensors on the primary and secondary reflectors permitted a dedicated measurement of the telescope emissivity at mission completion. The high sensitivity of the LFI radiometers, together with the optimal knowledge of LFI systematic effects, allowed the derivation of the telescope emissivity from the thermal excess measured by the LFI radiometers.\\ The emissivity measured is consistent with the \textit{on ground} Reflection Loss measured in the range 100 GHz - 380 GHz, extrapolated to the LFI frequencies. Slight deviations from the extrapolated curve are consistent with the improvements expected from the lower telescope temperature in flight. Extrapolation to 40K of '\textit{on ground}' reflection Loss measured at higher temperatures showed  that the telescope performance was not degraded at EOL w.r.t. BOL, and that the emissivity was about one order of magnitude better than the mission requirement.\\
The measure of success of future CMB experiments is how we cope with the knowledge of the systematic effects. Telescope emissivity can represent a large source of systematic uncertainties, since bigger and bigger mirrors will be required to feed thousands of receivers, needed to meet the ambitious requirements of next CMB experiments. This method could be hence usefully implemented for future space experiment at mm sub-mm wavelengths to finely characterize the telescope emissivity during the mission, in nominal conditions, provided that a dedicated thermal control system, based on control loop heaters and on a network of high resolution thermometers, is present.

\clearpage

\appendix
\section{Appendix}
\label{sec:Appendix}

\begin{table}[ht!]
  \begin{center}
    \caption{Calibration constants (K/V). Results are displayed per radiometer. M and S correspond to  \texttt{MAIN} and \texttt{SIDE} radiometers}.
    \label{cal_const}
    \begin{tabular}{l l l}
      \hline
      \hline
	&\multicolumn{2}{c}{R{\sc adiometer}} \\
	\multicolumn{1}{c}{RCA} & \multicolumn{1}{c}{M (K/V)}& \multicolumn{1}{c}{S (K/V)}\\
	\hline
\texttt{LFI18}&	14.2561	&	21.68022	\\
\texttt{LFI19}&	26.69584	&	41.77317	\\
\texttt{LFI20}&	25.07979	&	30.41089	\\
\texttt{LFI21}&	44.25306	&	41.59089	\\
\texttt{LFI22}&	62.7823	&	60.77838	\\
\texttt{LFI23}&	34.74583	&	51.49897	\\
\noalign{\vskip	8pt}			
\texttt{LFI24}&	287.79424	&	178.85588	\\
\texttt{LFI25}&	126.36797	&	126.56825	\\
\texttt{LFI26}&	170.71762	&	144.40973	\\
\noalign{\vskip	8pt}			
\texttt{LFI27}&	12.79919	&	15.27546	\\
\texttt{LFI28}&	15.81183	&	19.22634	\\
\hline
      \end{tabular}
\end{center}
\end{table}
\newpage

\begin{acknowledgements}
The Planck Collaboration acknowledges the support of: ESA; CNES, and CNRS/INSU-IN2P3-INP (France); ASI, CNR, and INAF (Italy); NASA and DoE (USA); STFC and UKSA (UK); CSIC, MICINN, and JA (Spain); Tekes, AoF, and CSC (Finland); DLR and MPG (Germany); CSA (Canada); DTU Space (Denmark); SER/SSO (Switzerland); RCN (Norway); SFI (Ireland); FCT/MCTES (Portugal); ERC and PRACE (EU). A description of the Planck Collaboration and a list of its members, indicating which technical or scientific activities they have been involved in, can be found at URL:\\
http://www.cosmos.esa.int/web/planck/planck-collaboration.\\
The Planck LFI project (including instrument development and operation, data processing and scientific analysis) is developed by an international consortium led by Italy and involving Canada, Finland, Germany, Norway, Spain, Switzerland, UK, and USA. The Italian contribution is funded by the Italian Space Agency (ASI) and INAF.\\

We want to give special thanks to the Planck Mission Operations Center (MOC) for the professionality and helpfulness shown during the  whole Planck mission and, with respect to this work, during the LFI EOL Test Campaign.\\
\end{acknowledgements}

\end{document}